# Visualizing plasmon-exciton polaritons at the nanoscale using electron microscopy

Andrew B. Yankovich[1], Battulga Munkhbat[1], Denis G. Baranov[1], Jorge Cuadra[1], Erik Olsén[1], Hugo Lourenço-Martins[2], Luiz H. G. Tizei[2], Mathieu Kociak[2], Eva Olsson[1,*], Timur Shegai[1,*]

1. *Department of Physics, Chalmers University of Technology, 412 96, Gothenburg, Sweden*

2. *Laboratoire de Physique des Solides, Univ. Paris-Sud, CNRS UMR 8502, Orsay, France*

\* eva.olsson@chalmers.se; \* timurs@chalmers.se

## Abstract

Polaritons are compositional light-matter quasiparticles that have recently enabled remarkable breakthroughs in quantum and nonlinear optics, as well as in material science [1–10]. Despite the enormous progress, however, a direct nanometer-scale visualization of polaritons has remained an open challenge [11]. Here, we demonstrate that plasmon-exciton polaritons, or plexcitons [12–14], generated by a hybrid system composed of an individual silver nanoparticle and a few-layer transition metal dichalcogenide can be spectroscopically mapped with nanometer spatial resolution using electron energy loss spectroscopy in a scanning transmission electron microscope. Our experiments reveal important insights about the coupling process, which have not been reported so far. These include nanoscale variation of Rabi splitting and plasmon-exciton detuning, as well as absorption-dominated extinction signals, which in turn provide the ultimate evidence for the plasmon-exciton hybridization in the strong coupling regime. These findings pioneer new possibilities for in-depth studies of polariton-related phenomena with nanometer spatial resolution.

# Main

Strong light-matter coupling occurs when the rate of coherent energy exchange between the optical cavity (light subpart) and quantum emitter (matter subpart) exceeds the system's constituents decay rates, resulting in vacuum Rabi splitting and the genesis of new hybrid quantum states called polaritons. Polaritons exhibit inseparable light and matter characteristics [15,16], leading to the modification of the system's chemical reactivity [9,10], charge transport [17], and potential energy landscapes [18]. Creating a strongly coupled system requires a careful engineering of both the optical cavity and the quantum emitter, so that they are in resonance and have optimal mode characteristics [15,16]. Traditionally, strong coupling has been realized using high finesse optical microcavities filled up with resonant absorbing materials, such as in solid-state [19] and atomic [20] systems, as well as by combining Fabry-Pérot microcavities [21] and nanoparticle diffractive arrays [22] with organic dyes. This led to several remarkable breakthroughs [1], including polariton lasing [2], photon blockade [3], Bose-Einstein condensation [4] and superfluidity [5]. However, the applicability of these systems can be limited by the low-temperature requirements, large device footprints or unstable molecules. More recently, localized surface plasmon (LSP) modes of metallic nanocavities allowed observation of plasmon-exciton polaritons, also known as plexcitons, thanks to their subwavelength mode volumes [12–14]. Plexcitons have enabled new research directions through demonstrations of strong coupling using individual molecules [6] and quantum dots [7], and room temperature polariton lasing [8]. In particular, plasmonic nanocavities strongly coupled to excitons in semiconducting mono- to few-layer transition metal dichalcogenide (TMDC) materials [23–25] enable small device footprints, room temperature operation, and more stable components.

To understand and realize the full potential of plexcitons, it is essential to have the ability to simultaneously spatially and spectrally resolve polaritons at their characteristic length and energy scales. Optical spectroscopy, with its impressive energy resolution, has played an integral role in advancing our knowledge of strong coupling; however, it is unable to image the subwavelength nature of polaritons because of its diffraction-limited spatial resolution. A scanning transmission electron microscope (STEM) allows for imaging materials with unprecedented spatial resolution down to the deep-sub-Angstrom scales [26]. When equipped with a modern electron monochromator and electron energy loss spectrometer (EELS), it also gains impressive spectroscopic capabilities [27]. The sensitivity of STEM EELS to optical and electronic excitations has allowed access to previously unattainable information about the fundamental behavior of LSPs in nanostructures [28–30] and excitons in TMDC materials [31]. A recent theoretical study has predicted that STEM EELS could be capable of detecting plexcitons [11], but this has not yet been experimentally validated.

Here, we experimentally demonstrate that STEM EELS can spectrally and spatially resolve strong light-matter coupling in plexciton systems. This is achieved by experimentally interrogating a hybrid material system composed of a single silver (Ag) truncated nanopyramid (TNP) coupled to a few-layer flake of tungsten disulfide ($WS_2$) (Fig. 1a), which has previously exhibited strong coupling signatures in optical dark field (DF) scattering [25]. To create the hybrid system, mono- and few-layer $WS_2$ flakes were mechanically exfoliated from high-quality bulk crystals and transferred onto ~20 nm thick $Si_3N_4$ membranes, followed by drop-casting of wet-chemically synthesized Ag TNPs (Methods, Supplementary Fig. 1). The flexibility of this system allows for correlative spectroscopy experiments using optical DF and EELS, and high spatially resolved imaging experiments using high angle annular dark field (HAADF) STEM imaging, on the same individual uncoupled and coupled systems.

The Ag TNPs used in this study have 70-110 nm side lengths, clear faceted surfaces, and sharp 5-10 nm radius corners (Fig. 1b). The TNPs are three-fold rotationally symmetric regular pyramids that are truncated at the top. Atomic resolution TEM images (Fig. 1d and Supplementary Fig. 2) reveal that the TNPs are single crystalline and have 0-3 nm shells composed of organic ligand remnants from the solution synthesis. Correlative optical DF and EELS experiments reveal the uncoupled TNPs host bright dipolar LSPs that support linewidths of about 200 meV (Fig. 1c). The dipolar LSP is excited with higher probability when the electron beam is positioned just outside the corners of the particle (Supplementary Fig. 3) and its resonance energy strongly depends on the TNP side length, truncation height, and corner sharpness, as well as the ligand shell thickness and the refractive index of any material in the vicinity of the particle [32]. The ~0.1 eV red-shift of the dipolar LSP when measured optically compared to EELS and the small variations in correlation agreement of different data sets can be attributed to experimental limitations that are discussed in the Methods. TNPs also host optically-dark higher-order LSPs (Supplementary Fig. 3) that are present in EEL spectra up to the bulk plasmon frequency; however, these modes are not involved in the strong coupling studied here. By design, the sizes of the Ag TNPs are chosen so that they host uncoupled dipolar LSPs that are slightly blue-shifted compared to the $WS_2$ A-exciton. Then, when creating the hybrid system, the high background permittivity of $WS_2$ [33] red-shifts the TNP dipolar LSP so that it spectrally overlaps with the $WS_2$ A-exciton, enabling their hybridization. When the two are perfectly overlapping in energy, the system is zero-detuned.

$WS_2$ flakes host Wannier-Mott excitons with ~50 meV linewidth [34]. Even though the $WS_2$ excitons have more than one order of magnitude smaller EELS excitation probability compared to the TNP dipole LSP, they are visible in experiments (Fig. 1c) because we use high signal-to-noise ratio experimental techniques [30] (Methods). The $WS_2$ A-exciton peak is observed at 2.03 eV in EELS, and blue shifts in both EELS and optical data as the $WS_2$ thickness is reduced to monolayer (Supplementary Fig. 4). Additionally, the uncoupled $WS_2$ B- and C-exciton peaks are observed in EELS at ~2.4 eV and ~2.75 eV [34], respectively, but these are not involved in the strong coupling studied here due to large detuning. The EELS exciton observations match simulated EELS predictions (Supplementary Fig. 4). The number of $WS_2$ layers in a flake can be independently determined from the HAADF STEM image intensity and the EELS signal (Methods, Supplementary Fig. 5).

It is important to begin this investigation with understanding how the EEL spectra from coupled systems evolve with the number of $WS_2$ layers and the instrument energy resolution. Figure 2 shows simulated EEL spectra from zero-detuned coupled systems as a function of the number of $WS_2$ layers (Methods). EELS simulations predict resolvable splitting for all $WS_2$ thicknesses by the presence of two distinct peaks, corresponding to the lower polariton (LP) and upper polariton (UP), with 80 and 110 meV peak splitting for monolayers and 6 layers, respectively (Fig. 2a). The increase in splitting from mono- to multilayers is caused by the Ag TNP LSP interacting with more excitons when positioned on top of a thicker $WS_2$ flake [24]. The splitting magnitude is predicted to saturate at ~120 meV with >8 $WS_2$ layers (Supplementary Fig. 6). However, in practice, EELS resolution has instrumental limitations (see Methods). Therefore, when the EELS energy resolution is reduced to 70 meV (Fig. 2b), an upper bound in our experiments, the simulated EEL spectra exhibit less resolved splitting. For monolayer $WS_2$ the splitting is no longer resolvable, while for 6 layers it still is. Nevertheless, when the 70 meV resolution simulated data were fitted using the coupled mode theory (Methods), the extracted plasmon-exciton coupling strengths ($g$) were identical to those extracted from the not broadened simulated data (Supplementary Fig. 6).

Figure 2b also shows two experimental EEL spectra from coupled systems containing monolayer and 6-layers $WS_2$ flakes. For both coupled systems, the TNP dipole LSP red-shifts to ~2 eV and overlaps with the A-exciton line in the $WS_2$ flake. More importantly, for the 6 layers $WS_2$ coupled system, the dipole LSP peak splits into two distinct peaks separated by ~130 meV with a dip located near the A-exciton resonance. However, the splitting disappears in the EELS measurement from the monolayer $WS_2$ coupled system, agreeing with the simulated predictions. For this reason, the remainder of the study focuses on coupled systems containing 6 layers $WS_2$ flakes.

To corroborate that the system is in the strong coupling regime and the observed splitting is indeed vacuum Rabi splitting, we first correlate the optical DF and EEL spectra and demonstrate anti-crossing of the LP and UP polaritons (Fig. 3). Fig. 3a shows correlated optical DF and EEL spectra from the same coupled system, revealing good agreement of the observed splitting. Correlative data from other coupled systems show similar peak splitting but with varying correlative agreement (Supplementary Fig. 7) because of the previously discussed experimental limitations (Methods). The typical method of showing plexciton anti-crossing is to controllably vary the LSP energy using one geometrical parameter and scan it through the exciton resonance. This is extremely challenging in our system because there are multiple parameters varying from particle to particle affecting the LSP energy as previously discussed. Instead, we experimentally show plexciton anti-crossing by measuring multiple coupled systems and arranging them by their relative detuning (Fig. 3b). By comparing the HAADF STEM images (Fig. 3b) and the experimental spectrum, it is evident that other characteristics than only the TNP side length affect the degree of detuning. The experimental EELS anti-crossing behavior agrees with simulated EELS behavior, which was calculated by only varying the TNP model side length (Fig. 3c). Additionally, clear anti-crossing is seen in simulated optical scattering, TNP absorption, total absorption and extinction signals (Fig. 3d-g), which were calculated using the same models as the EELS anti-crossing simulations. The absorption within the TMDC exhibits signs of anti-crossing (Fig. 3h) indicating that the system is at the onset of strong coupling [35]. It is important to note that EELS probes extinction (total loss) by the coupled system, rather than its scattering (radiative loss) [36]. Previous observations of Rabi splitting in individual plasmonic nanoparticles were experimentally confirmed by light scattering techniques, which are subject to criticism because there is potential for other far-field interference effects resembling Rabi splitting[16]. Therefore, observation of mode splitting in EELS provides stronger evidence for the realization of vacuum Rabi splitting than splitting in optical DF scattering spectra[35].

We further inspect the strength of the light-matter interaction by directly extracting the coupling strength, $g$, from the experimental EEL spectra in Fig. 3b using the coupled mode theory (Methods). This analysis yields $g$ in the range of 53 to 71 meV depending on the system (Fig. 3b). Comparing $g$ to the exciton linewidth ($\gamma_X \sim 70$ meV) and the LSP linewidth ($\gamma_{pl} \sim 210$ meV), which are extracted from the optical spectra in order to avoid artificial EELS broadening, one can see that the condition for mode hybridization $g > \frac{|\gamma_{pl} - \gamma_X|}{4}$ holds, indicating the occurrence of Rabi oscillations in the system [15,16] (Supplementary Fig. 8). Additionally, the Rabi frequency, which can be calculated as $\Omega = 2\sqrt{g^2 - \frac{(\gamma_{pl} - \gamma_X)^2}{16}}$, is closely approaching the $\frac{\gamma_{pl} + \gamma_X}{2}$ limit, thus indicating the onset of strong coupling [15,16]. We emphasize that these extracted Rabi frequency values are consistently lower than the observed peak-to-peak distance in EEL and DF spectra, which closely follow twice the coupling strength $2g$ and thus overestimate the actual Rabi splitting. To gain additional evidence for the plasmon-exciton hybridization in this system, we performed several control experiments, where the $WS_2$ multilayer was displaced from the Ag TNP (Supplementary Fig. 9). This data shows no evidence of splitting, thus proving that simple additive absorption in Ag TNP and $WS_2$ multilayer cannot explain the observations in Fig. 3.

Finally, we use STEM EELS to spatially map plexcitons and reveal their sub-optical-wavelength confinement. Fig. 4a shows EEL spectra from the three corners of a coupled system that exhibits clear Rabi splitting and is near the zero-detuned condition. However, these spectra also reveal that this particular coupled system generates hybridized plexcitons with different characteristics depending on which TNP corner is probed. These nanoscale-resolved differences are evident from the variations in the plasmon-exciton detuning energies, $\omega_{pl} - \omega_X$, that were extracted using the coupled mode theory (Fig. 4a). Fig. 4b-e are experimental STEM EELS maps extracted from the same spectrum image that was used to create the spectra in Fig. 4a at the LP (1.94 eV) and UP (2.07 eV) peak maxima (for maps of higher order modes see Supplementary Fig. 10). The maps in Fig. 4b-c were extracted from the non-normalized spectrum image and reflect the as-acquired LP and UP signals. These maps reveal the LP and UP exhibiting clear localization at the TNP corners, but they contain effects from other electron scattering sources than just the LP and UP, such as Ag elastic diffraction (Methods). To counteract some of these effects, each spectrum in the image was normalized to its zero-loss peak intensity. Figures 4d-e show maps that more closely reflect the EEL from only the LP and UP because they are extracted from the normalized spectrum image. These maps also reveal clear localization of the LP and UP at the TNP corners. Simulated EELS maps (Methods) of the LP and UP of a coupled system that was modeled after the experimentally measured system (Fig. 4f-g), show excellent agreement with the experimental normalized maps. The better agreement with the normalized data is expected because the simulations predict only the effects of electron scattering from the coupled systems optical density of states. In addition, simulated electric field distributions (Fig. 4h-i) and simulated surface charge density distributions (Fig. 4j-k) reveal that both the LP and UP display a behavior that is characteristic of the dipolar LSP (Supplementary Fig. 3). These EELS maps indicate that the electric field distributions of the plexcitons are completely driven by its dipolar LSP.

To conclude, we have demonstrated that vacuum Rabi splitting from strongly coupled plexcitons is experimentally resolvable at their fundamental energy and length scales by STEM EELS. Because STEM EELS measures an absorption-dominated extinction signal, it provides strong evidence for the plasmon-exciton hybridization than the typically used optical DF scattering techniques [35]. By utilizing the strong coupling between LSPs hosted by Ag TNPs and excitons hosted in few-layer $WS_2$ flakes, we have observed up to 130 meV splitting in EEL spectra that correlates well with optical measurements and agrees with theoretical predictions. Experimental EELS observations of anti-crossing between the LP and UP, and the analysis of the extracted coupling strength, support the conclusion that this system is at the onset of the strong coupling regime. Additionally, we demonstrate that STEM EELS allows for spatially mapping the LP and UP with nanometer resolution, which reveals that the plexcitons adopt a plasmon-like charge and field distribution. They also may have different characteristics depending on the local nanoscale environment and where the coupled system is probed with the electron beam. Our results establish STEM EELS as an essential characterization tool for improving the understanding of plasmon-exciton interactions and strong coupling. In a broader perspective, because STEM EELS combines unprecedented spatial resolution, extremely wide spectral range (~100 meV – 1000 eV), and impressive spectral resolution (down to ~5 meV), we anticipate it will enable new possibilities for in-depth studies of other polariton-related phenomena and systems in the near future.

**Figures:**

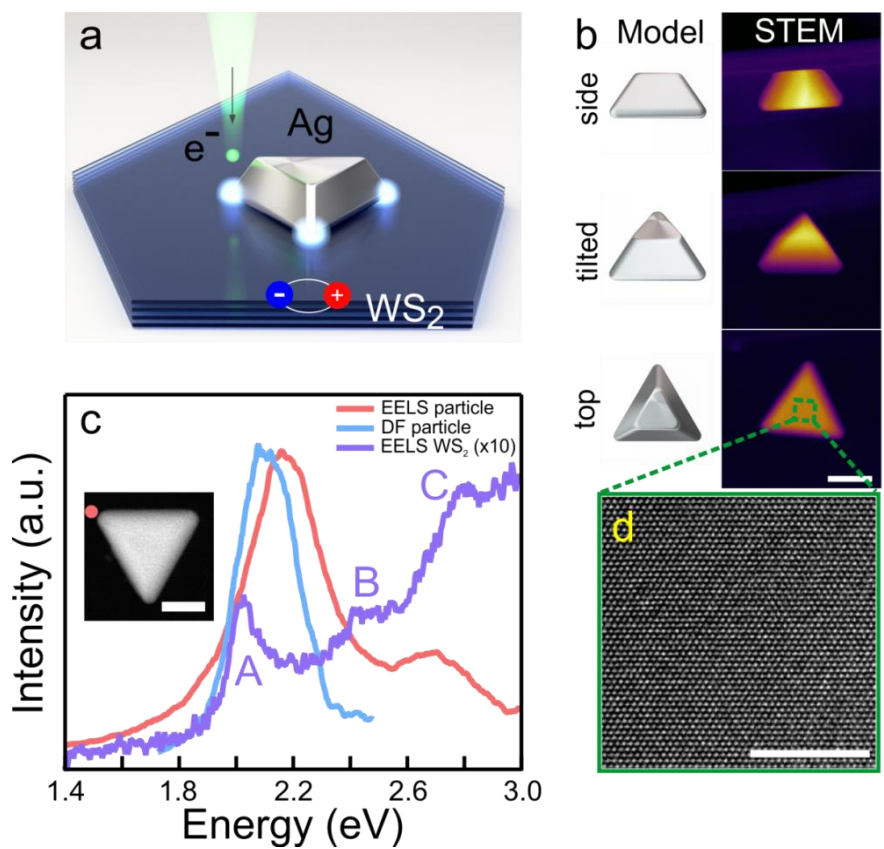

**Fig. 1: The components of the plasmon-exciton hybrid system.** (a) Schematic diagram showing an electron beam exciting a strongly coupled PEP system composed of a Ag TNP and few-layer $WS_2$. (b) Tilt-series HAADF STEM images of the same TNP supported on a lacey carbon TEM grid, along with an inferred model in similar projection orientations. (c) Correlated optical DF scattering spectra (blue) and EEL spectra (red) from the same 105 nm uncoupled Ag TNP that is shown in the inset HAADF STEM image. The EEL spectrum was acquired at the corner marked by the red circle. Also included is an EEL spectrum (purple, multiplied by 10 to increase visibility) from an uncoupled 6 layers $WS_2$ flake showing signals from the A, B, and C-excitons. (d) Atomic resolution TEM image of a typical TNP, such as the area marked by the white box in (b). The scale bars in (b), (c) and (d) are 50, 50 and 5 nm, respectively.

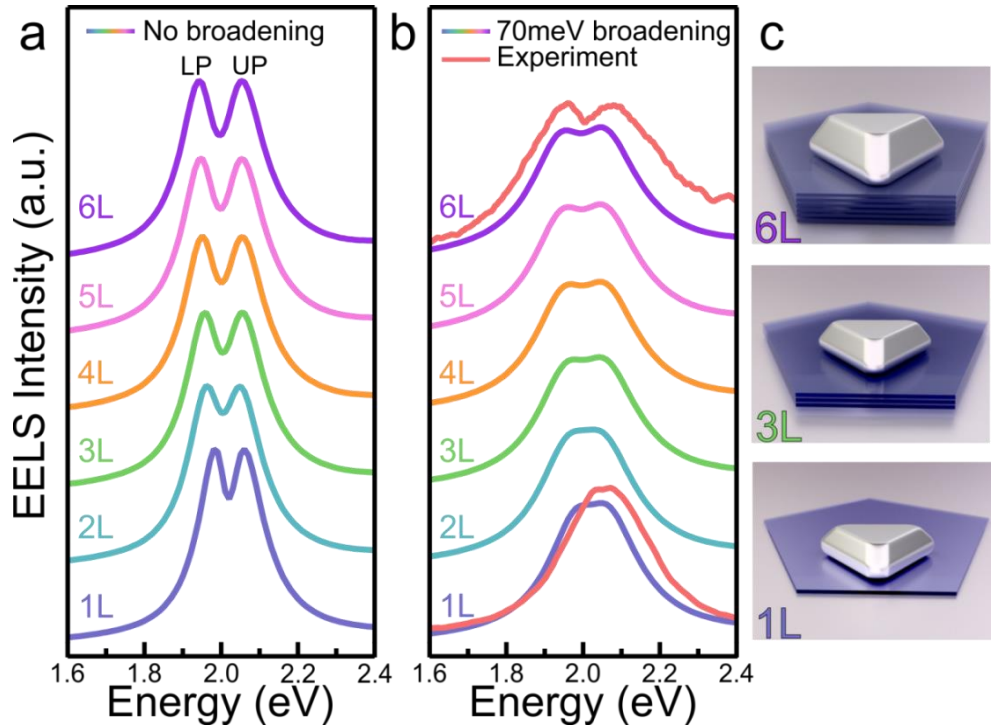

**Fig. 2: The effect of WS$_2$ thickness and EELS energy resolution on Rabi splitting observations.** (a) and (b) Calculated EEL spectra from the corner of a perfectly tuned coupled system that is composed of a 75 nm Ag TNP on a WS$_2$ flake on a 20 nm Si$_3$N$_4$ support. The WS$_2$ flake thickness was varied between 1 and 6 layers. The perfectly tuned situation, where the LP and UP have the same peak intensities, was found by varying the height of each TNP while keeping the particle length constant. (a) Assumes no spectral broadening from the experimental limitations, while (b) assumes an energy resolution of 70 meV, which is an upper limit for our 40-70 meV resolution experiments. The red spectra in (b) are experimental EEL spectra from coupled TNPs on 1 and 6 layers WS$_2$. (c) Schematics of the models used to achieve the zero-detuned condition for 1, 3, and 6 WS$_2$ layers. Notice that in order to achieve perfect tuning, the TNP height was changed for different WS$_2$ thicknesses.

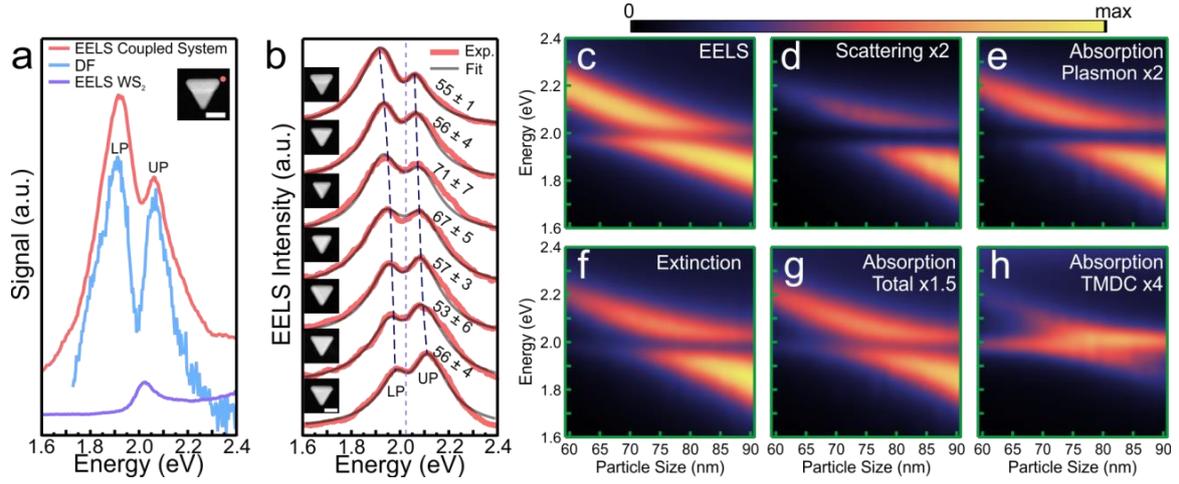

**Fig. 3: Experimental and theoretical polariton anti-crossing.** (a) Correlated experimental optical DF scattering spectrum (blue) and EEL spectrum (red) from the coupled TNP – 6 layers $WS_2$ system shown in the inset HAADF STEM image. The red EEL spectrum was acquired at the corner marked by the red circle. The purple EEL spectrum is acquired for the electron beam positioned on the $WS_2$ a few hundred nanometers away from the TNP. (b) Experimental EEL spectra (red) from seven TNPs on the same 6 layers $WS_2$ flake with various degrees of detuning. The dashed purple line shows the position of the $WS_2$ A-exciton and the dashed black lines show the shifts of the LP and UP peak positions. The top spectrum has a red-shifted LSP compared to the A-exciton, the middle spectrum is near the zero detuned condition, and the bottom spectrum has a blue-shifted LSP. HAADF STEM images of each TNP are shown in the inset images. The scale bars in (a) and (b) are 50 nm, and the scale bar in (b) is applicable to all images in (b). The grey lines are the fits of the experimental EEL spectra using the coupled mode theory. The plasmon-exciton coupling strength ($g$, meV) for each coupled system, which was extracted through the fitting, is reported beside each spectrum. (c-h) Calculated EEL and optical spectra as a function of TNP side length, revealing anti-crossing behavior in (c) EELS, (d) optical scattering, (e) optical absorption from the TNP, (f) optical extinction, (g) optical total absorption, and (h) optical absorption from the TMDC. EELS and optical simulations utilized the same models. (c) has its own intensity scale, while (d-h) are all on the same intensity scale but with some signals multiplied by the factor indicated in the figure to make all the signals visible.

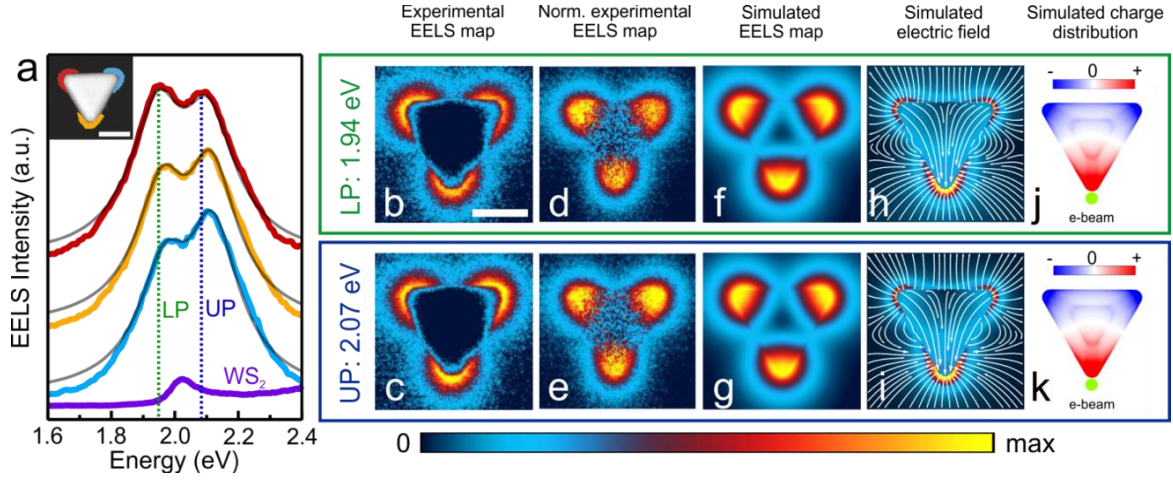

**Fig. 4: Spatially and spectrally resolving PEPs using STEM EELS.** (a) The red, yellow, and blue EEL spectra are from each corner of the coupled TNP – 6 layers WS$_2$ system that is shown in the HAADF STEM image (inset). The spectra were extracted from a spectrum image within the colored overlay regions of the HAADF STEM image. The LP and UP peaks in the red spectra are marked by the vertical dotted lines at 1.94 eV and 2.07 eV, respectively. The grey lines are the fits of the experimental EEL spectra using the coupled mode theory. The plasmon-exciton coupling strengths $g$, extracted through fitting, are 74±10, 70±11 and 67±14 meV for the red, yellow, and blue spectra, respectively. The plasmon–exciton detuning energies ($\omega_{pl} - \omega_X$), extracted through fitting, are -6±3, 14±3 and 29±5 meV for the red, yellow, and blue spectra, respectively. The purple EEL spectrum is from the uncoupled WS$_2$ flake far away from the TNP. (b) and (c) are EELS maps of the coupled system shown in (a) extracted at the LP and UP energies, respectively, from a spectrum image that has not been normalized. (d) and (e) are EELS maps extracted at the LP and UP energies, respectively, from the spectrum image that has been normalized to the zero-loss peak height. (f) and (g) are simulated EELS maps extracted at the LP and UP energies respectively. (h) and (i) are FDTD simulated electric field distributions in the plane at the bottom of the TNP at the LP and UP energies, respectively, excited by a vertically polarized and normally incident plane wave. (j) and (k) are simulated surface charge density distributions at the LP and UP energies, respectively, excited by an electron beam passing along the green marker. The scale bars in (a) and (b) are 50 nm.

## Acknowledgements

We acknowledge the Swedish Research Council (VR) and the Engkvist Foundation for financial support. This project has received funding from the European Union's Horizon 2020 research and innovation program under grant agreement No 823717 – ESTEEM3 and the National Agency for Research under the program of future investment TEMPOS-CHROMATEM with the Reference No. ANR-10-EQPX-50. This work was performed in part at the Chalmers Material Analysis Laboratory, CMAL.